\documentclass[preprint]{aastex}

\shorttitle{Constraining the Fe Abundances in Nebulae}
\shortauthors{Rodr\'\i guez \& Rubin}

\begin{document}

\title{The [\ion{Fe}{4}] Discrepancy: Constraining the Iron Abundances in
	Nebulae}

\author{M\'onica Rodr\'\i guez}
\affil{Instituto Nacional de Astrof\'\i sica, \'Optica y Electr\'onica, INAOE,\\
Apdo Postal 51 y 216, 72000 Puebla, Pue., Mexico}
\email{mrodri@inaoep.mx}

\author{Robert H. Rubin\altaffilmark{1}}
\affil{NASA Ames Research Center, Moffett Field, CA 94035--1000}
\altaffiltext{1}{Orion Enterprises}
\email{rubin@cygnus.arc.nasa.gov}

\begin{abstract}
We study the current discrepancy between the model-predicted and measured
concentrations of Fe$^{++}$ and Fe$^{3+}$ in ionized nebulae. We calculate a
set of photoionization models, updated with the atomic data relevant to the
problem, and compare their results with those derived for the available nebulae
where both [\ion{Fe}{3}] and [\ion{Fe}{4}] lines have been measured. Our new
model results are closer to the measured values than the results of previous
calculations, but a discrepancy remains. This discrepancy translates into an
uncertainty in the derived Fe abundances of a factor up to $\sim4$. We explore
the possible causes of this discrepancy and find that errors in the Fe atomic
data may be the most likely explanation. The discrepancy can be fully accounted
for by any of the following changes: (1) an increase by a factor of $\sim10$ in
the recombination rate (radiative plus dielectronic, or charge
transfer) for Fe$^{3+}$, (2) an
increase by a factor of 2--3 in the effective collision strengths for
Fe$^{++}$, or (3) a decrease by a factor of 2--3 in the effective collision
strengths for Fe$^{3+}$. We derive the Fe abundances implied by these three
explanations and use the results to constrain the degree of depletion of Fe in
our sample nebulae. The Galactic \ion{H}{2} regions and planetary nebulae
are found to have high depletion factors, with less than 5\% of
their Fe atoms in the gas phase. The extragalactic \ion{H}{2} regions
(LMC~30~Doradus, SMC~N88A, and SBS~0335$-$052) have somewhat lower depletions.
The metal-deficient blue compact galaxy SBS~0335$-$052 could have from 13\% to
40\% of Fe in the gas phase. The depletions derived for the different objects
define a trend of increasing depletion at higher metallicities.
\end{abstract}

\keywords{\ion{H}{2} regions --- ISM: abundances}

\section{INTRODUCTION}

The high depletion factors found for Fe in the interstellar medium (ISM), down
to $\log(\mbox{Fe}/\mbox{H})-\log(\mbox{Fe}/\mbox{H})_\odot=-2.3$
\citep{sav96}, and the relatively high cosmic abundance of this element, imply
that Fe is a very important contributor to the mass of refractory dust grains
\citep*{sof94}.
The high depletion factors also imply that the destruction of a small quantity
of dust grains will translate into a significant, i.e. measurable, increase of
the Fe abundance in the gaseous phase.
Hence, the study of the Fe abundance in the gas of different regions where
different conditions prevail can be used to identify the processes that govern
the evolution of dust in the ISM.\@
In the diffuse ISM, the depletion patterns found for all
available elements, including Fe, have led to the identification of shock waves
as the main destruction mechanism of dust (\citealt{jen04} and references
therein).
In \ion{H}{2} regions, the lack of strong lines from other refractory elements
and the reasons outlined above imply that Fe is the best choice to study
depletion trends. Such a study, based on the Fe abundances measured in several
Galactic \ion{H}{2} regions, suggests that energetic photons are
responsible for the destruction of some dust grains in these nebulae
\citep{rod96,rod02}.

Fe is expected to be in three ionization states in \ion{H}{2} regions:
$\mbox{Fe}/\mbox{H}=\mbox{Fe}^{+}/\mbox{H}^{+}+
\mbox{Fe}^{++}/\mbox{H}^{+}+\mbox{Fe}^{3+}/\mbox{H}^{+}$, so that the
measurement of \ion{Fe}{2}, \ion{Fe}{3}, and \ion{Fe}{4} emission lines will
allow us to determine the Fe abundance in these nebulae.
[\ion{Fe}{2}] and [\ion{Fe}{3}] lines, although weak, have already been
observed in several \ion{H}{2} regions and starburst galaxies (e.g.\@
\citealt{izo99,rod02}).
Most of the optical [\ion{Fe}{2}] lines are affected by fluorescence effects
\citep{rod99,ver00}, but the Fe$^{+}$ abundance can be estimated from a few
lines that are almost insensitive to fluorescence. The Fe$^{+}$ abundance
turns out to be low in most of the \ion{H}{2} regions
studied to date \citep{rod02}, as expected from the low ionization
potential for this ion (16.2 eV).
On the other hand, [\ion{Fe}{4}] lines are much weaker than [\ion{Fe}{2}] and
[\ion{Fe}{3}] lines and hence very difficult to observe.
Therefore, the Fe abundance in \ion{H}{2} regions is usually obtained from
[\ion{Fe}{3}] lines and an ionization-correction factor ($ICF$), derived
from photoionization models, to account for the contribution of Fe$^{3+}$.
The relation
\begin{equation}
\label{eq1}
\frac{\mbox{Fe}}{\mbox{O}}=ICF\,\frac{\mbox{Fe}^{++}}{\mbox{O}^+}=
\frac{x(\mbox{O}^+)}{x(\mbox{Fe}^{++})}\frac{\mbox{Fe}^{++}}{\mbox{O}^+},
\end{equation}
where $x(X^{n+})$ stands for the ionization fraction of the $X^{n+}$ ion, is
especially well suited for determining the Fe abundance from optical
observations of \ion{H}{2} regions, since (1) the ionization potentials of the
O and Fe ions are similar (30.6 and 54.8 eV for Fe$^{++}$ and Fe$^{3+}$, 35.3
and 54.9 eV for O$^{+}$ and O$^{++}$), and (2) both O$^{+}$ and O$^{++}$ can be
measured from strong optical lines and one can get the O abundance
$\mbox{O}/\mbox{H}=\mbox{O}^{+}/\mbox{H}^{+} + \mbox{O}^{++}/\mbox{H}^{+}$, and
hence also $\mbox{Fe}/\mbox{H}$ from $\mbox{Fe}/\mbox{O}$. The values of the
above $ICF$ and their dependence, if any, with the degree of ionization can be
found using grids of photoionization models. The available grids of models
\citep{sta90,gru92} imply that a constant value for this $ICF$,
$x(\mbox{O}^{+})/x(\mbox{Fe}^{++})=1.1$, should provide a good estimate of the
total Fe abundance to within $\pm0.2$~dex.

However, available measurements of some weak [\ion{Fe}{4}] lines
\citep{rub97,rod03} imply Fe$^{3+}$ abundances that are smaller, by factors
3--8, than the values implied by equation (\ref{eq1}) with an $ICF$ equal
to 1.1.
This ``[\ion{Fe}{4}]  discrepancy'' translates into an uncertainty of up to a
factor of 5 in the Fe abundances derived for a wide range of objects, from the
nearby Orion Nebula to the low metallicity blue compact galaxy SBS~0335$-$052.
Thus, the discrepancy has important implications for our understanding of the
evolution of dust in \ion{H}{2} regions, the dependence of dust depletion
factors on metallicity, and the chemical history of low metallicity dwarf
galaxies.

In this paper we study the ionization equilibrium of Fe using
photoionization models that incorporate recently improved values for all the
atomic data relevant to the problem.
We compare the new model results with the available observational data, discuss
the possible reasons behind the [\ion{Fe}{4}] discrepancy, and study its effect
on the reliability of the derived Fe abundances.

\section{MODEL RESULTS}

We have used the photoionization code NEBULA (see, e.g, \citealt{rub91a,rub91b}
and references therein)
to calculate a grid of spherically symmetric models of constant density ionized
by a single star.
We use this grid to determine the value of
$x(\mbox{O}^+)/x(\mbox{Fe}^{++})$, the $ICF$ in equation (\ref{eq1}), and
its dependence on the degree of ionization given by
$x(\mbox{O}^+)/x(\mbox{O}^{++})=\mbox{O}^+/\mbox{O}^{++}$.

We have updated NEBULA with the atomic data derived recently from improved
calculations that are relevant to the problem: photoionization cross sections
for Fe$^{+}$, Fe$^{++}$, O$^{0}$ and O$^{+}$ \citep{nah94,nah96a,kj02b,ver96},
recombination coefficients for Fe$^{++}$, Fe$^{3+}$, O$^{+}$ and O$^{++}$
\citep{nah96b,nah97,nah99}, the charge-exchange reactions involving these ions
(\citealt{kin96}; and the ORNL/UGA Charge Transfer Database for
Astrophysics\footnote{http://www-cfadc.phy.ornl.gov/astro/ps/data/home});
and the NLTE model stellar atmospheres of \citet*{stern03} for solar
metallicity with surface gravity $\log(g)=4$.
The photoionization cross section of Fe$^{+}$ was constructed using both the
calculated values of \citet{nah94} and the experimental ones of \citet{kj02b},
following the prescriptions given by the later authors.
Other recent upgrades to the code are described by \citet{sim04}.

Our grid of 36 models covers the following parameter space: effective
temperature of the ionizing star $T_{\rm eff}=35\,000$, $40\,000$,
$45\,000$, and 50\,000~K; total nucleon density $N=100$,
1000, and $10\,000$~cm$^{-3}$; and ``Orion metallicity'' Z$_{\rm
Orion}$ ($\mbox{He}/\mbox{H}=0.1$, $\mbox{C}/\mbox{H}=3.3\times10^{-4}$,
$\mbox{N}/\mbox{H}=4.5\times10^{-5}$, $\mbox{O}/\mbox{H}=4.0\times10^{-4}$,
$\mbox{Ne}/\mbox{H}=8.1\times10^{-5}$, $\mbox{S}/\mbox{H}=2.2\times10^{-5}$,
$\mbox{Ar}/\mbox{H}=4.5\times10^{-6}$, $\mbox{Si}/\mbox{H}=3.0\times10^{-6}$,
$\mbox{Fe}/\mbox{H}=3.0\times10^{-6}$), Z$_{\rm Orion}/10$, and Z$_{\rm
Orion}/30$.

All the fluxes of the ionizing stars were normalized to get a total
number of ionizing photons s$^{-1}$ for hydrogen of $10^{49}$, but we checked
that this has no effect on the $ICF$:
we ran two new models with $10^{51}$ ionizing photons s$^{-1}$ and found that
they follow the same trend of $x(\mbox{O}^+)/x(\mbox{Fe}^{++})$ versus
$\mbox{O}^+/\mbox{O}^{++}$ defined by the original grid.
This same consistent behavior was found when we used one of the supergiant
models of \citet{stern03}, with $T_{\rm eff}=40\,000$~K and $\log(g)=3.4$, as
the ionizing star.

The photoionization cross sections for the O and, especially, the Fe ions show
some sharply peaked resonances arising from excitations to autoionizing states
(quasi-bound states, above the ionization threshold). 
The energies at which these resonances occur, their widths and their peak
intensities can be uncertain by a few percent or more, as seen in the direct
comparison with experimental data in the few instances where the later are
available (e.g.\@ \citealt{kj02a,kj02b} for O$^+$ and Fe$^+$, respectively).
For this reason, and for reasons of computational ease, the photoionization
cross sections are usually smoothed or fitted with simple functions.
We used the analytic fits of \citet{ver96} for O$^{0}$ and O$^{+}$, and,
following \citet*{baum98}, we smoothed the photoionization cross sections for
Fe$^{+}$ and Fe$^{++}$ doing a convolution with a Gaussian of width 3\% the
energy.
The fluxes of the stellar atmospheres were also smoothed by convolving with a
Gaussian of width 1\% the energy.

\begin{figure}
\plotone{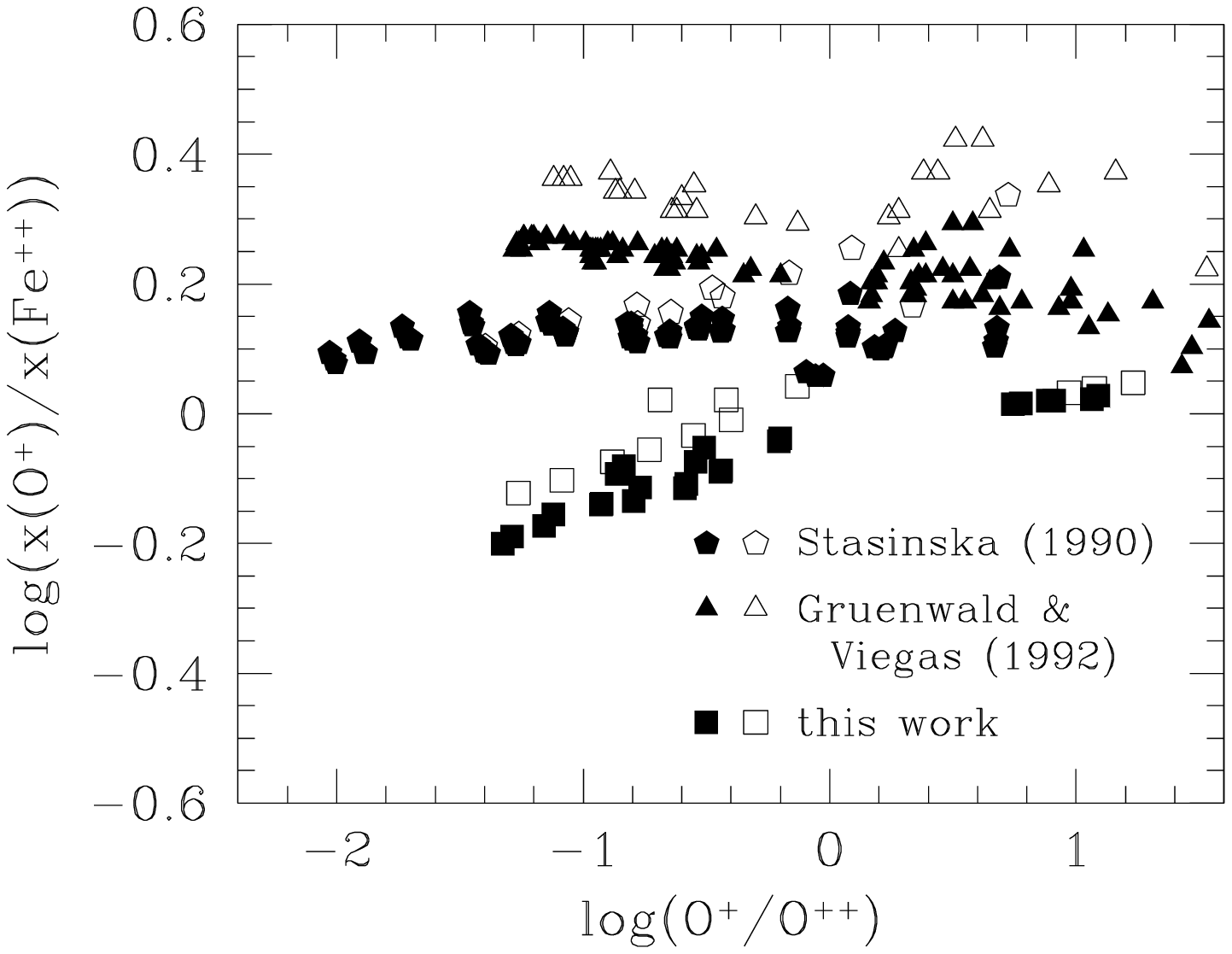}
\caption{Our new model results for $x(\mbox{O}^{+})/x(\mbox{Fe}^{++})$ as a
function of $\mbox{O}^{+}/\mbox{O}^{++}$ and the previous results by
\citet{sta90} and \citet{gru92}. {\sl Open symbols:} results for solar or
near solar metallicity (Z$_{\rm Orion}$).
{\sl Filled symbols:} results for lower metallicities, from $Z_\odot/2$ to
$Z_\odot/50$.}
\label{sgv}
\end{figure}

Figure~\ref{sgv} shows the values of the $ICF$ in equation (\ref{eq1}),
$x(\mbox{O}^{+})/x(\mbox{Fe}^{++})$, obtained from various models as a
function of the degree of ionization given by $\mbox{O}^{+}/\mbox{O}^{++}$.
The results of previous ionization models \citep{sta90,gru92} for metallicities
that go from solar to 1/50 of solar, are also shown for comparison.
Our new models show lower values for the $ICF$, and a small dependence with the
degree of ionization.
It can also be seen in Figure~\ref{sgv} that the results for solar or near solar
metallicity show slightly larger values for the $ICF$ than the results for
lower metallicities. This is due to
the relatively high optical depth reached in the outer parts of these
solar-metallicity models at
the O$^+$ ionization edge. At lower metallicities this optical depth becomes
negligible. This dependence on the metallicity is small and will not be
further considered.
A least-squares fit to the new model results in Figure~\ref{sgv} leads to the
ionization-correction scheme:
\begin{equation}
\label{eq2}
\frac{\mbox{Fe}}{\mbox{O}}=
\frac{x(\mbox{O}^+)}{x(\mbox{Fe}^{++})}\frac{\mbox{Fe}^{++}}{\mbox{O}^+}=
0.9\,\biggl(\frac{\mbox{O}^{+}}{\mbox{O}^{++}}\biggr)^{0.08}\,
	\frac{\mbox{Fe}^{++}}{\mbox{O}^+}.
\end{equation}

\section{COMPARISON WITH THE OBSERVATIONS}

The $ICF$ implied by the models can be compared with the values derived
empirically for the handful of objects where the observed spectra include any
[\ion{Fe}{4}] line (along with diagnostic lines, [\ion{O}{2}],
[\ion{O}{3}], and [\ion{Fe}{3}] lines).
To the objects considered by \citet{rod03}, we have added several objects where
[\ion{Fe}{4}]~$\lambda6739.8$ has been observed recently: the \ion{H}{2}
regions M42 \citep{est04} and NGC~3576 \citep{gar04}, and three of the
planetary nebulae (PNe) observed by \citet{liu04a}: NGC~6210, NGC~6826,
and NGC~6884.
\citet{liu04a} provided the intensities of other [\ion{Fe}{4}] lines in some
objects, but all of them are blends or possible misidentifications.

Physical conditions and ionic abundances have been derived following the same
procedure outlined in \citet{rod03}, except that the new values for the
Fe$^{3+}$ transition probabilities of \citet{fro04} have been used for all the
objects.
The Fe$^{3+}$ abundances implied by these new data differ from the values
presented by \citet{rod03} by less than 20\%.
Table~\ref{t1} shows the physical conditions used in the abundance
determination;
Table~\ref{t2} shows the ionic and total abundances derived for all the
objects.
In all the objects but two (N88A and SBS~0335$-$052), the abundances have been
derived with the usual two-zone scheme: we used the electron temperature
implied by the [\ion{O}{3}] diagnostic lines, $T_e$[\ion{O}{3}], to derive the
O$^{++}$ and Fe$^{3+}$ abundances, and $T_e$[\ion{N}{2}] to derive the O$^+$
and Fe$^{++}$ abundances.
In N88A and SBS~0335$-$052, we used $T_e$[\ion{O}{3}] to derive all ionic
abundances.
We could have used instead an uncertain estimate of $T_e$[\ion{N}{2}] obtained
from $T_e$[\ion{O}{3}] and one of the existing relations between the two
$T_e$'s (either empirical or derived from photoionization models), but this
would not change our results in a significant way.
For example, with the relation of \citet*{cam86} we obtain
$T_e\mbox{[\ion{N}{2}]}=12\,900$~K in N88A~bar.
If we had used this $T_e$ instead of $T_e\mbox{[\ion{O}{3}]}=14\,200$~K to
derive the O$^+$ and Fe$^{++}$ abundances in this object, the total Fe and O
abundances presented in Table~\ref{t2} would change by less than 0.05~dex, and
the value of any of the ionic ratios we will be considering would remain within
the error bars.

As in \citet{rod03}, it has been assumed that $\mbox{O}/\mbox{H}=
\mbox{O}^+/\mbox{H}^++\mbox{O}^{++}/\mbox{H}^+$, $\mbox{Fe}/\mbox{H}=
\mbox{Fe}^{+}/\mbox{H}^++\mbox{Fe}^{++}/\mbox{H}^++\mbox{Fe}^{3+}/\mbox{H}^+$,
and that the Fe$^+$ abundance is negligible in those objects showing a high
degree of ionization.
The spectra of the PNe and SBS~0335$-$052 show some lines from high ionization
ions, such as [\ion{O}{4}], [\ion{Fe}{5}], [\ion{Fe}{6}] or [\ion{Fe}{7}],
but we expect that only traces of these ions are likely to be present in our
sample objects.
\citet{liu04b} calculated $\mbox{O}^{3+}/\mbox{H}^+\sim3\times10^{-5}$ from the
[\ion{O}{4}] line at $25.9$~$\mu$m measured by the Infrared Space Observatory
({\sl ISO}) in NGC~6884.
This $\mbox{O}^{3+}$ abundance is just 8\% of our adopted O abundance.
\citeauthor{liu04b} also estimated that the contribution of $\mbox{O}^{3+}$
to the total abundance would be $\sim10$ times lower for NGC~6210 and completely
negligible for NGC~6826.
Furthermore, since $\mbox{Fe}^{3+}$ has an ionization potential very close to
that of $\mbox{O}^{++}$ ($54.8$ and $54.9$~eV, respectively), $\mbox{Fe}^{4+}$
is not likely to have a significant concentration.
Similar considerations, based on the $\mbox{He}^{++}/\mbox{He}^+$ ratio, were
used by \citet{rod03} to conclude that the concentrations of $\mbox{O}^{3+}$ and
$\mbox{Fe}^{4+}$ are likely to be very low for the other high ionization
objects.

\begin{figure}
\plotone{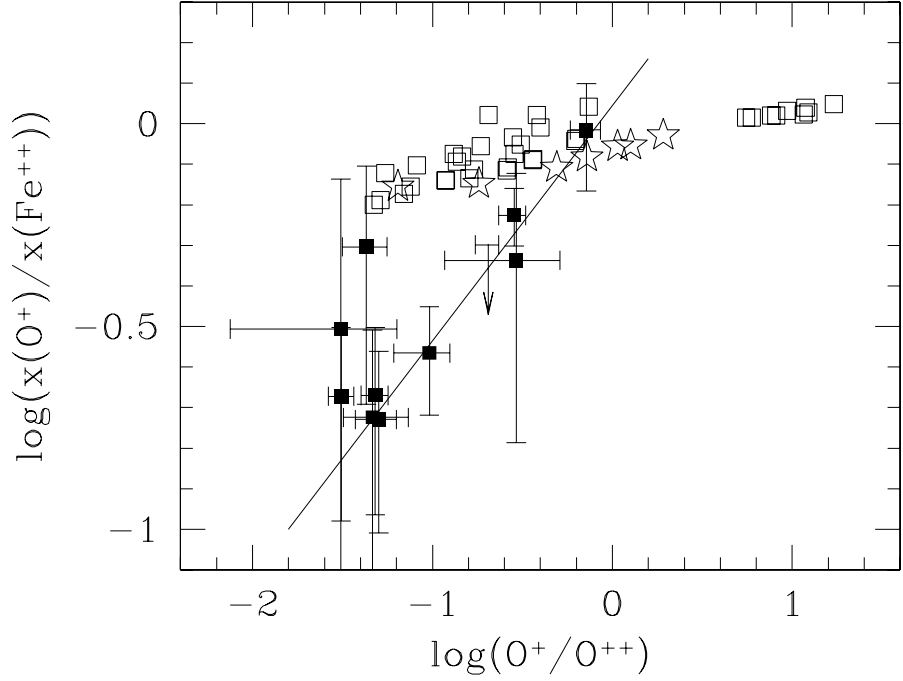}
\caption{Values of $x(\mbox{O}^{+})/x(\mbox{Fe}^{++})$ as a function of
$\mbox{O}^{+}/\mbox{O}^{++}$ for our new models and for the observed objects.
{\sl Open squares:} Results for the spherical models. {\sl Stars:} Results for
the shell models. {\sl Filled squares:} Values calculated for
the observed objects. From left to right: IC~4846 (ordinate $-0.51$),
NGC~6826, NGC~6210, N88A square A, SBS~0335$-$052, NGC~6884, N88A bar,
30~Doradus (the upper limit), M42~b, M42~a, and NGC~3576. The line shows a
least-squares fit to the data for those objects with
$\log(\mbox{O}^{+}/\mbox{O}^{++})>-1.35$.}
\label{ru}
\end{figure}

The values of $x(\mbox{O}^{+})/x(\mbox{Fe}^{++})$ versus
$\mbox{O}^{+}/\mbox{O}^{++}$ implied by the results in Table~\ref{t2} are
compared with the model results in Figure~\ref{ru}.
The new model results, although closer to the measured values than were the
previous model predictions, are still unable to explain the measured values.
We constructed additional models with a different geometry,
where the ionized gas is located in a shell around the star.
We used Z$_{\rm Orion}/10$, $T_{\rm eff}=45\,000$ and $50\,000$~K,
$N=100$, 1000, and $10\,000$~cm$^{-3}$, and internal radii
for the shell in the range 0.1--5.4 pc.
Typical results, shown in Figure~\ref{ru} as stars, are very close to those
obtained from the spherical models.
We also calculated the $ICF$ implied by different lines of sight through the
spherical model with Z$_{\rm Orion}/10$, $T_{\rm eff}=45\,000$~K, and
$N=100$~cm$^{-3}$.
The results followed the same trend defined by the shell models, and did not
help in explaining the discrepancy.
Hence, we did not pursue this approach further.

One could also speculate that the discrepancy is due to the fact that we are
comparing simple constant-density models with complex real objects.
\citet*{moo04} found that this can lead to significant errors in ratios like
$\mbox{O}^{+}/\mbox{O}^{++}$.
However, according to the models, Fe$^{++}$ and O$^+$ should have
similar concentrations, and [\ion{O}{2}] and [\ion{Fe}{3}] lines should form in
similar regions in the nebula, and hence these ions should be affected by
similar systematic effects.

\citet{rod03} discussed the most likely explanations of this discrepancy,
namely, errors in the collision strengths used to derive the Fe$^{++}$ and
Fe$^{3+}$ abundances or errors in the $ICF$ derived from models, probably
arising from errors in the input parameters governing the ionization
equilibrium.

If the model-predicted $ICF$ is seriously wrong, then the trend defined in
Figure~\ref{ru} by the observed objects will lead to an $ICF$ that should
be more reliable than the one predicted by the model results.
The trend is only clearly defined for those objects whose degree
of ionization is within the range covered by the models
(i.e. with $\log(\mbox{O}^{+}/\mbox{O}^{++})$ above $\sim-1.35$).
We note that since our photoionization models are tailored for \ion{H}{2}
regions, the highest $T_{\rm eff}$ we are considering is $50\,000$~K.\@
For this $T_{\rm eff}$, there are very few ionizing photons with
energies above $\sim54$~eV and O$^{++}$ and Fe$^{3+}$ (both with ionization
potentials above 54~eV) are not expected to be substantially further ionized.
The three objects with higher degree of
ionization in Figure~\ref{ru} are PNe, where the central stars
can reach or surpass $T_{\rm eff}$ of $100\,000$~K.
Even if these objects do not have significant concentrations of either O$^{3+}$
or Fe$^{4+}$, as discussed above, they might have small amounts of these ions
that could lead to a change in the trend followed by the $ICF$ with
$\log(\mbox{O}^{+}/\mbox{O}^{++})$.
Hence, a change in the trend at $\log(\mbox{O}^{+}/\mbox{O}^{++})\sim-1.4$ does
not seem unlikely.
Thus we limit the relationship to define an $ICF$ for lower degrees of
ionization.
A least-squares linear fit to the data for those objects with
$\log(\mbox{O}^{+}/\mbox{O}^{++})>-1.35$ in Figure~\ref{ru} leads to the
ionization-correction scheme:
\begin{equation}
\label{eq3}
\frac{\mbox{Fe}}{\mbox{O}}=
\frac{x(\mbox{O}^+)}{x(\mbox{Fe}^{++})}\frac{\mbox{Fe}^{++}}{\mbox{O}^+}=
1.1\,\biggl(\frac{\mbox{O}^{+}}{\mbox{O}^{++}}\biggr)^{0.58}\,
	\frac{\mbox{Fe}^{++}}{\mbox{O}^+},
\end{equation}
which should be valid for $-1.35<\log(\mbox{O}^{+}/\mbox{O}^{++})<-0.1$.
For $\log(\mbox{O}^{+}/\mbox{O}^{++})\geq-0.1$, the concentrations of Fe$^{++}$
and O$^+$ will grow, making these ions the dominant ionization states, and a
constant $ICF$:
\begin{equation}
\label{eq4}
\frac{\mbox{Fe}}{\mbox{O}}=\frac{\mbox{Fe}^{+}+\mbox{Fe}^{++}}{\mbox{O}^+},
\end{equation}
would be the preferred choice. The contribution of Fe$^+$ will still be very
small for most objects \citep{rod02}.
In \S~\ref{why} we consider what changes in the relevant input parameters that
affect the ionization equilibrium of the models would be needed to explain the
discrepancy.
Changes in the values of the collision strengths are considered in \S~\ref{col}.

\section{AN ERROR IN THE MODELS' $ICF$?
\label{why}}

The $ICF$ implied by the photoionization models depends on the following
factors: (1) the number of photoionizations of Fe$^{++}$ and O$^{+}$,
which in turn depends on the photoionization cross sections for these ions and
on the spectral energy distribution of the ionizing flux, (2) the number of
radiative and dielectronic recombinations of Fe$^{3+}$ and O$^{++}$, which
depends on the recombination coefficients, and (3) the rate of the
charge-exchange reactions leading to recombinations of Fe$^{3+}$ and O$^{++}$.

The smoothing of the photoionization cross sections and the stellar atmospheres
could introduce errors in the number of photoionizations computed in the
models. In order to constrain these errors, we calculated the number of
photoionizations of Fe$^{++}$ implied by the original stellar fluxes and
photoionization cross section of Fe$^{++}$ by integrating the product of these
two quantities between $h\nu_0=30.65$~eV, the ionization potential of Fe$^{++}$,
and $h\nu_1=54.4$~eV, the ionization potential of He$^+$ and the maximum energy
considered in the photoionization models.
We compared the result with the number of
photoionizations implied by the smoothed stellar fluxes and found that the
original value was 1\% lower than the smoothed one for $T_{\rm
eff}=35\,000$~K and 18, 6, and 9\% higher for $T_{\rm eff}=40\,000$,
$45\,000$, and 50\,000~K, respectively. These differences are far too small
to change our results in a significant way.

\begin{figure}
\plotone{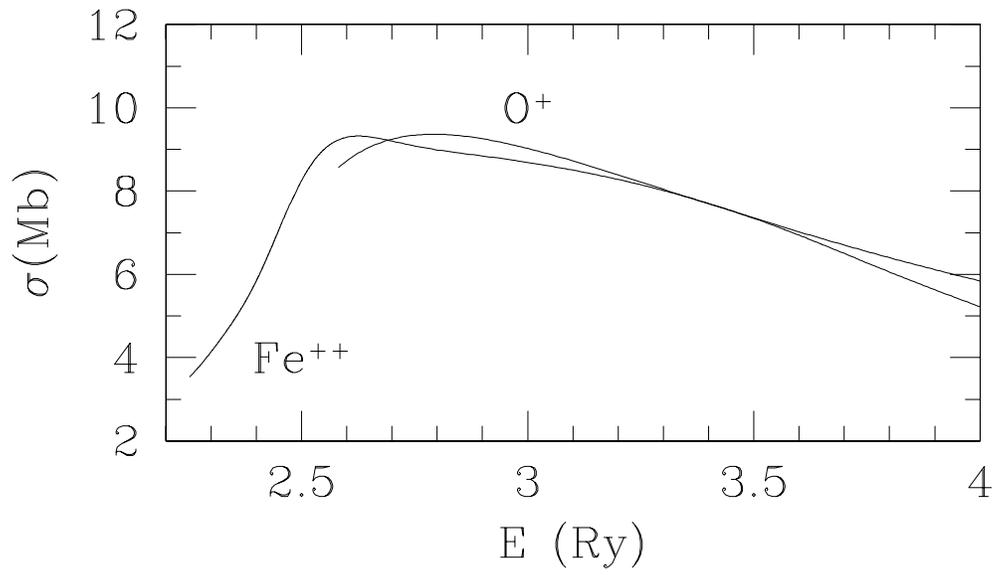}
\caption{Smoothed ionization cross sections for O$^+$ and Fe$^{++}$ as a
function of energy. Note that they are very similar in the energy range they
have in common.}
\label{ip}
\end{figure}

\begin{figure}
\plotone{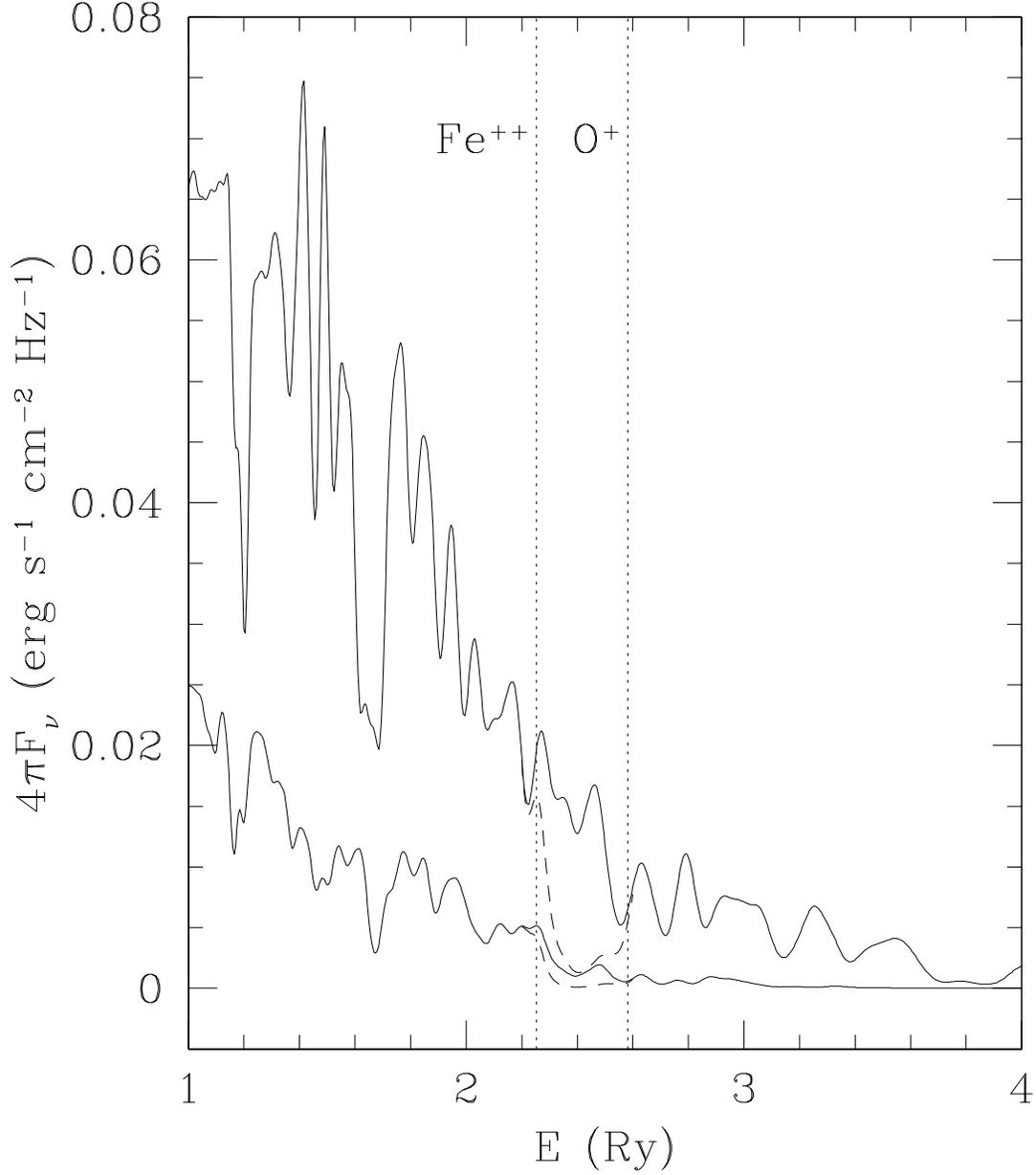}
\caption{{\sl Solid lines:} smoothed model stellar atmosphere spectral energy
distributions for $T_{\rm eff}=40\,000$~K (lower curve) and $50\,000$~K
(upper curve).
{\sl Dashed lines:} modified test section of the
spectra that we used to check the effect
of a lower ionizing flux in the region delimited by the ionization thresholds
of O$^+$ and Fe$^{++}$ (dotted lines). See the text for further information.}
\label{at}
\end{figure}

We consider now the effect of changes in the stellar ionizing flux.
The ionizing fluxes could be wrong because of uncertainties in the
stellar atmosphere models, or because we are using the results for models with
solar metallicity whereas lower metallicities might be more appropriate. 
We can constrain the kind of changes we need to check by noting that
the smoothed photoionization cross sections for Fe$^{++}$ and O$^{+}$ are very
similar for energies above the O$^{+}$ ionization threshold (Fig.~\ref{ip}).
Therefore, only a change in the ionizing flux in the energy range between the
two ionization thresholds (i.e. between $30.65$ and $35.12$ eV) can have a
significant effect on the $ICF$. A lower ionizing flux in this energy range
will change the $ICF$ in the right direction to solve the discrepancy.
We did a test with two of the model atmospheres, dividing their fluxes by
factors of up to a factor of ten in the energy interval of interest (see
Fig.~\ref{at}), thus bringing the fluxes in this interval very close to zero.
We achieved this by dividing the fluxes by the function
$1+ 9 \exp[-0.5((E-2.145)/0.06)^2]$, where E is the energy in Ry.
We ran two models using these new stellar fluxes (with $N=100$~cm$^{-3}$ and
Z$_{\rm Orion}$), and found that the decrease in the $ICF$ was $\simeq0.07$
dex, far too small to explain the discrepancy. We believe that this rules out
any uncertainty in the stellar ionizing flux distributions as the main cause
behind the discrepancy.

We are then left with errors in the atomic data governing the ionization
equilibrium of O and Fe as the possible explanations of the discrepancy.
Since the Fe ions are more complex than the O ions, their atomic data are more
difficult to calculate and hence more uncertain. We will center our
discussion on the effects of changes in the ionization and recombination data
for Fe.
Changes in the data for O going in the opposite direction from those we will
consider for the Fe data would also help in explaining the discrepancy, but we
note that the discrepancy was first discovered as related to Fe by considering
tailor-made models for M42, i.e., the discrepancy seems to be independent of the
degree of ionization given by the O ions \citep{rub97}.
We consider the $\mbox{Fe}^{++}\longleftrightarrow\mbox{Fe}^{3+}$ ionization
equilibrium, since Fe$^+$ has a low concentration in all the objects of our
sample. At a given point in a nebula, the ionization-equilibrium equation for
these two ions is given by (see, e.g.\@ \citealt{ost89}):
\begin{equation}
\label{eq5}
N(\mbox{Fe}^{++})\int_{\nu_0}^{\nu_1}\frac{4\pi J_\nu}{h\nu}\,
	\sigma(\mbox{Fe}^{++})\,d\nu
=N(\mbox{Fe}^{3+})N_e\,\alpha(\mbox{Fe}^{++},T_e)+
N(\mbox{Fe}^{3+})\,N(\mbox{H}^0)\,\delta(T_e),
\end{equation}
where $N(X)$ is the volume density of $X$, $J_\nu$ is the mean intensity of the
radiation field at the point, $\sigma(\mbox{Fe}^{++})$ is the photoionization
cross section for Fe$^{++}$, $\alpha(\mbox{Fe}^{++},T_e)$ is the total
(radiative plus dielectronic) recombination coefficient of $\mbox{Fe}^{3+}$,
and $\delta(T_e)$ is the rate coefficient of the charge-exchange reaction
$\mbox{Fe}^{3+} + \mbox{H}^0 \rightarrow \mbox{Fe}^{++} + \mbox{H}^+$.

\begin{figure}
\plotone{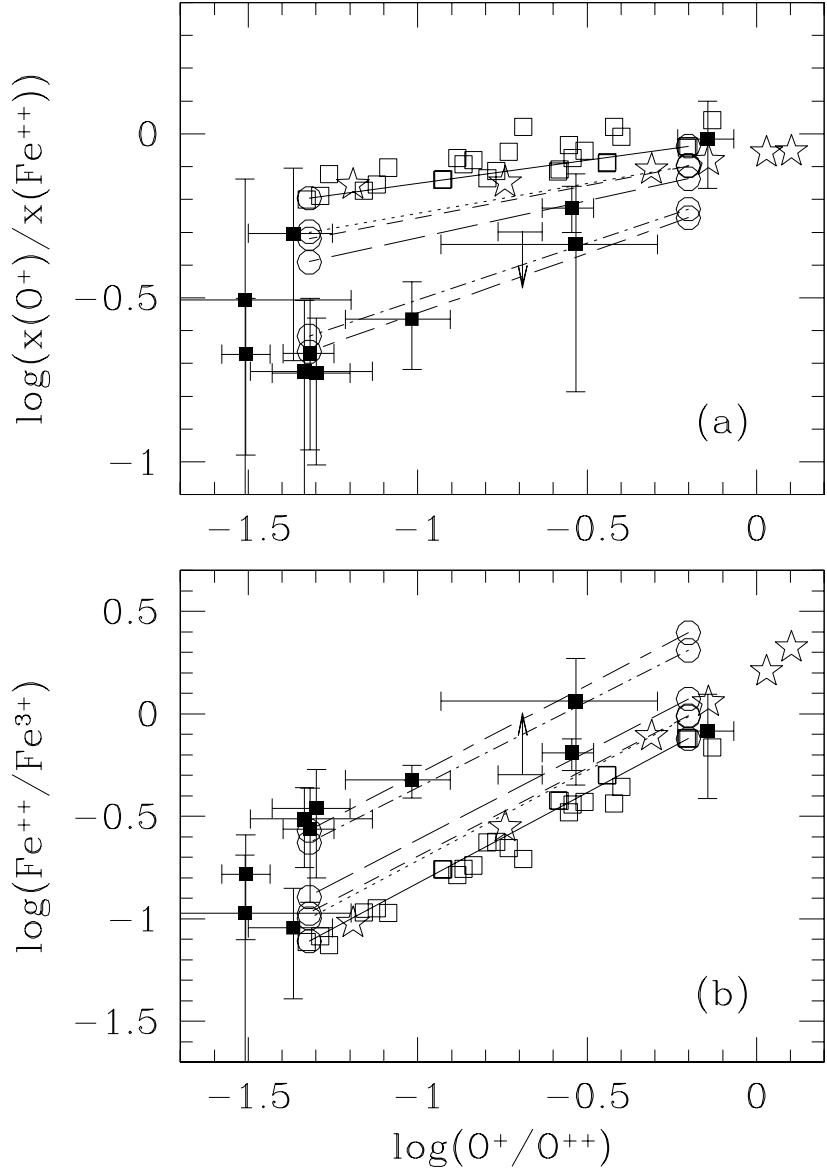}
\caption{Values of $x(\mbox{O}^{+})/x(\mbox{Fe}^{++})$ {\sl (a)} and
$\mbox{Fe}^{++}/\mbox{Fe}^{3+}$ {\sl (b)} as a function of
$\mbox{O}^{+}/\mbox{O}^{++}$ for our new models (open squares and stars,
see Fig.~\ref{ru} for more information) and for the observed objects
(filled squares).
{\sl Connected open circles:} Predictions from the test models described in
\S\ref{why}.
From top to bottom in panel {\sl (a)} and from bottom to top in panel
{\sl (b)}:  (1) the two models used as templates (connected by a solid line),
(2) models with the Fe$^{3+}$ recombination coefficient increased by a factor
of 2 (dotted line),
(3) models with the rate of the charge exchange reaction
$\mbox{Fe}^{3+} + \mbox{H}^0 \rightarrow \mbox{Fe}^{++} + \mbox{H}^+$ increased
by a factor of 2 (short dashed line),
(4) models with the Fe$^{++}$ ionization cross-section decreased by a factor of 2
(long dashed line),
(5) models with the rate of the charge exchange reaction increased by a factor
of 10 (dot-dashed line),
and (6) models with the Fe$^{3+}$ recombination coefficient increased by a
factor of 10 (short-long dashed line).}
\label{cha}
\end{figure}

For a given $\mbox{O}^{+}/\mbox{O}^{++}$ we can get a lower value of the
$ICF$
$x(\mbox{O}^{+})/x(\mbox{Fe}^{++})$ by decreasing the photoionization cross
section for Fe$^{++}$ or increasing either the recombination coefficient of
Fe$^{3+}$ or the rate of the aforementioned charge-exchange reaction.
We selected two models to use as templates. They are ionized by stars with
$T_{\rm eff}=40\,000$~K and $10^{49}$ ionizing photons s$^{-1}$, and
$T_{\rm eff}=50\,000$~K, $10^{51}$ ionizing photons s$^{-1}$; both have
$N=100$~cm$^{-3}$ and Z$_{\rm Orion}/10$.
We then tested sequentially the effect on the results of these models of
changes by a factor of 2 in the photoionization cross section, and by factors
of 2 and 10 in each of the recombination coefficient and the rate of the
charge-exchange reaction.
We did not consider a change by a factor of 10 in the photoionization cross
section of Fe$^{++}$ because the comparisons of calculated cross sections with
the available experimental data (e.g.\@ \citealt{kj02a,kj02b} for O$^+$ and
Fe$^+$, respectively) usually show better agreement.
The results from the two model templates and from the test calculations are
shown as connected open circles in Figure~\ref{cha}, where it can be
seen that a change by a factor of 10 in the recombination data will be needed
in order to reproduce the observed results.
A factor of 10 uncertainty would not be unexpected for the dielectronic part of
a recombination coefficient or for the rate of a charge-exchange reaction
\citep{fer03}.

\section{ERRORS IN THE COLLISION STRENGTHS FOR Fe$^{++}$ OR Fe$^{3+}$?
\label{col}}

\begin{figure}
\plotone{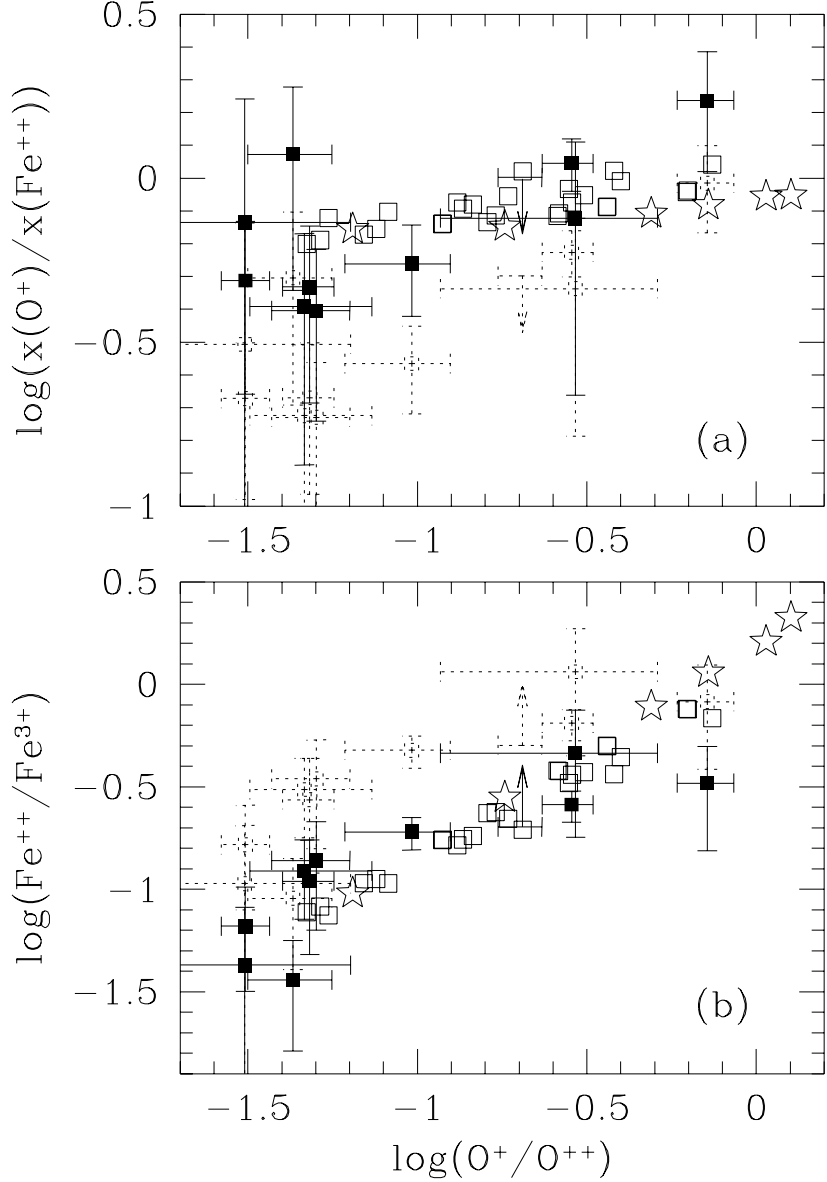}
\caption{Values of $x(\mbox{O}^{+})/x(\mbox{Fe}^{++})$ {\sl (a)} and
$\mbox{Fe}^{++}/\mbox{Fe}^{3+}$ {\sl (b)} as a function of
$\mbox{O}^{+}/\mbox{O}^{++}$ for our new models (open squares and stars,
see Fig.~\ref{ru} for more information) and for the observed objects
if the Fe$^{3+}$ abundance is multiplied by a factor of 2.5 or the Fe$^{++}$
abundance is divided by the same factor (filled squares). Dotted lines show the
original positions of the observed objects.}
\label{cham}
\end{figure}

Figure~\ref{cham}a shows the comparison between the
$x(\mbox{O}^{+})/x(\mbox{Fe}^{++})$ $ICF$ implied by models
and observations when either the derived Fe$^{++}$ abundances are divided by a
factor $2.5$ or the Fe$^{3+}$ abundances are multiplied by the same factor.
Figure~\ref{cham}b shows the same comparison for $\mbox{Fe}^{++}/\mbox{Fe}^{3+}$
as a function of $\mbox{O}^{+}/\mbox{O}^{++}$.
It can be seen that this factor of 2.5 change in the relative abundances of
Fe$^{++}$ and Fe$^{3+}$ would lead to an agreement between observations and
models.
It might then look promising that the recent calculations of collision
strengths for Fe$^{++}$ by \citet{mcl02} differ from the previous results we
are using here \citep{zha96} by factors up to 2.
This new atomic data might then imply Fe$^{++}$ abundances
lower by a factor of 2, thus reducing the discrepancy in Fig.~\ref{ru}.
However, \citet{mcl02} calculate the collision strengths only for transitions
between terms, whereas the fine-structure values are needed to derive Fe$^{++}$
abundances and, most important in the present context, to check their
reliability by comparing the predicted relative line intensities with the
observed ones.
The atomic data we are using here for [\ion{Fe}{3}] do seem to be reliable,
since they lead to consistent abundances for the various [\ion{Fe}{3}] lines
(\citealt{rod02,rod03}) and, when the [\ion{Fe}{3}] lines are used as an
electron density diagnostic, they lead to values similar to those implied by
the usual diagnostics such as [\ion{S}{2}], [\ion{Cl}{3}] or [\ion{Ar}{4}]
\citep{gar04,est04}.
However, it is possible to preserve this consistency while changing the values
of the collision strengths so that they lead to lower Fe$^{++}$ abundances.
Since the upper levels of the optical [\ion{Fe}{3}] lines are mainly populated
through collisional transitions from the ground term, the relative intensities
of the lines we are considering will not change significantly if all the
collision strengths for transitions originating in the ground term are changed
by a similar factor.
It is therefore suggestive that all the term-averaged collision strengths of
\citet{zha96} for transitions from the ground term are lower than the results
given by \citet{mcl02} by factors $\sim2$.
A test calculation shows that if the collision strengths of \citet{zha96} for
transitions from the ground term are enhanced by a factor of 2, the
[\ion{Fe}{3}] line ratios remain mostly unaffected and the Fe$^{++}$ abundances
are lower by a factor of $\sim2$.
Hence, this change in the Fe$^{++}$ collisional data would explain most, if not
all, the discrepancy.
New calculations that provide the collision strengths for the fine-structure
levels will be extremely valuable in order to test this idea.

On the other hand, the discrepancy might be due to errors in
the atomic data for Fe$^{3+}$, which are difficult to test through a comparison
between observed and predicted relative line intensities
because the lines are very weak and difficult to measure.
Our new model results along with the new observational results for M42 and,
especially, for NGC~3576 in Figure~\ref{ru}, which are very close to the
expected ones, allow us to rule out the large uncertainties in the collision
strengths, of factors 6--7, contemplated by \citet{rod03}, and to settle for
changes by factors 2--3.

A comparison between the results in Figures~\ref{cha} and \ref{cham} shows that
both the explanation involving errors in the collision strengths of Fe$^{3+}$
or Fe$^{++}$ by a factor 2--3 and the explanation requiring a change in the
recombination data for Fe$^{3+}$ (the recombination coefficient or the
rate of the charge-exchange reaction with H$^0$) by a
factor of 10 look equally plausible (it must be taken into account that the
model results show some dispersion around their defined trends).
Of course, the final explanation for the discrepancy might require a
combination of causes, but the fact that we cannot decide on the most likely or
on the more important one,
introduces an uncertainty in the Fe abundances calculated for the various
objects.
In the next section, we assess this uncertainty and try to see what constraints
we can place on the Fe abundance in the nebulae of our sample.

\section{CONSTRAINING THE IRON ABUNDANCES}

The last two columns in Table~\ref{t2} show the Fe abundances derived for all
the objects from the sum of the ionic abundances (col.~[8]) and from the
Fe$^{++}$ abundance and the $ICF$ (see eq.~[\ref{eq2}]) implied by our
photoionization models (col.~[9]).
If the model predicted $ICF$ is seriously wrong, the best values for the Fe
abundance will be those shown in column~(8); if the discrepancy is only due to
errors in the Fe$^{3+}$ collision strengths, the best values will be those in
column~(9); and if the discrepancy is mainly due to errors in the Fe$^{++}$
collision strengths by the factor of $\sim2$ suggested by the calculations of
\citet{mcl02} (see \S~\ref{col}), the best values will be those shown in
column~(9) lowered by $\sim0.3$~dex.
Figure~\ref{dep} shows the depletion factors for the Fe/O abundance ratio
($[\mbox{Fe}/\mbox{O}]=\log(\mbox{Fe}/\mbox{O})-\log(\mbox{Fe}/\mbox{O})_\odot$)
implied by these three possibilities as a function of the O/H abundance ratio.
We note that if the discrepancy is due to some combination of the
aforementioned causes, the errors required in any of the atomic data are likely
to be lower than those considered above, and the depletion factors will
consequently be intermediate between those shown in the three panels of
Figure~\ref{dep}.
We use Fe/O to calculate depletion factors because our objects have different
metallicities (see the values of the O/H abundance ratio in Table~\ref{t2}) but
are likely to have a near solar value for Fe/O (considering the abundances in
gas and dust), or at least, their intrinsic Fe/O abundance will show less
variation than either O/H or Fe/H.
Indeed, the Fe/O abundance ratio has been found to be near solar or slightly
($\sim0.2$~dex) below solar for the
Magellanic Clouds and for other low-metallicity dwarf galaxies (see e.g.,
\citealt{ven01}; \citet*{she01}, and references therein).
We have used $\log(\mbox{Fe}/\mbox{O})_\odot=-1.2$, a value that agrees to
within $\pm0.1$~dex with recent determinations of the solar Fe and O abundances
\citep{hol01,asp04,mel04}.

Even if we do not know which are the correct values for the nebular Fe
abundances, several things can be inferred from the results in Figure~\ref{dep}.
Since we are considering quite extreme variations of the atomic
data entering in the abundance determinations, we can use the range of
abundances as a reasonable constraint on the true Fe abundances.
An inspection of the depletion factors in Figure~\ref{dep} shows that the
Galactic \ion{H}{2} regions (M42, NGC~3576) and PNe (IC~4846,
NGC~6210, NGC~6826, and NGC~6884) have depletion factors in the range $-1.3$ to
$-2.0$, intermediate between the values observed for warm and cold clouds in the
Galactic ISM, where typical depletion factors are $\sim-1.2$ and $\sim-2.2$,
respectively \citep{sav96}.
Thus most of the Fe atoms are in dust grains in these nebulae:
less than $\sim5$\% of their total Fe abundance is present in the gas.
The depletion factor in LMC~30~Doradus, where the metallicity is about $0.2$~dex
lower, is at the higher end of the above range, $\sim-1.4$.
In SMC~N88A, with about $0.5$~dex lower metallicity than in the Galactic
objects, the depletion factor is lower, in the range $-0.5$ to $-1.1$.
The blue compact galaxy SBS~0335$-$052, with a metallicity $-1.3$~dex below
those of the Galactic nebulae, shows a somewhat lower depletion, in the range
$-0.4$ to $-0.9$.
The depletions could be somewhat lower for the metal-poor objects if
their intrinsic Fe/O abundance ratios are below solar, as commented above.
This trend of higher depletions at higher metallicities, which was also shown to
hold for a smaller sample of objects by \citet{rod02}, is consistent with what
we know about depletion factors in the ISM of the Magellanic
Clouds \citep{wel99,wel01} and with a recent measurement of the gas to dust
ratio in the SMC \citep{bot04}.
A similar trend has been found to hold for Damped Ly$\alpha$ systems
\citep{vla04}.
The trend could arise from a low efficiency of dust-formation processes at low
metallicities or, at least for the objects in our sample, from a high
dust-destruction rate due to the harsh radiation fields usually found in
metal-poor galaxies.
Further measurements of both [\ion{Fe}{3}] and [\ion{Fe}{4}] lines in a sample
of metal-poor galaxies would help to constrain this issue.

We note that the depleted Fe atoms are not likely to be in the form of silicates
in N88A.\@
\citet*{roc87} did not find the $9.7$~$\mu$m silicate feature in the IR spectrum
of this \ion{H}{2} region, and \citet{kurt99} derived a solar value for the Si/O
abundance ratio.
Furthermore, \citet{wel01} found that Si (and Mg) are essentially undepleted in
the SMC ISM, concluding that silicates cannot be an important
component of dust in this galaxy.
This lack of silicate dust does not seem to be a common feature of
low-metallicity galaxies, since the $9.7$~$\mu$m feature has been detected in
SBS~0335$-$052 \citep{hou04}.
The depleted Fe may be in the form of oxides or metallic grains, or deposited
onto carbon grains.
Such Fe-containing non-silicate dust grains are also considered to be an
important dust component in our own Galaxy, where less than half of the
depleted Fe can be accounted for with silicates (see \citealt{whi03} and
references therein).

\section{SUMMARY AND CONCLUSIONS}

We have presented a detailed analysis of the current discrepancy between the
observationally derived and model predicted concentrations of Fe ions in
ionized nebulae.
We have calculated new photoionization models that incorporate state-of-the-art
values for all the atomic data relevant to the problem.
The predicted Fe ionic concentrations have been compared with those implied by
the available observational data.
Our new model results are closer to the observed values than previous
calculations, but there is still a discrepancy that translates into an
uncertainty in the derived Fe abundances of a factor up to $3.7$.
We have studied the possible reasons for this discrepancy and conclude that the
most likely explanations are those due to uncertainties in the atomic data.
We are able to find a satisfactory agreement between the model predictions and
the observations in three different ways: (1) increasing either the total
recombination coefficient or the rate of the charge exchange reaction with H$^0$
for Fe$^{3+}$ by a factor of $\sim10$, (2) decreasing the collision strengths
for Fe$^{3+}$ by a factor of 2--3, and (3) increasing the collision
strengths for Fe$^{++}$ by a factor of 2--3.
Of course, if errors in different atomic data are involved, the above factors
need not be as large.

\begin{figure}
\plotone{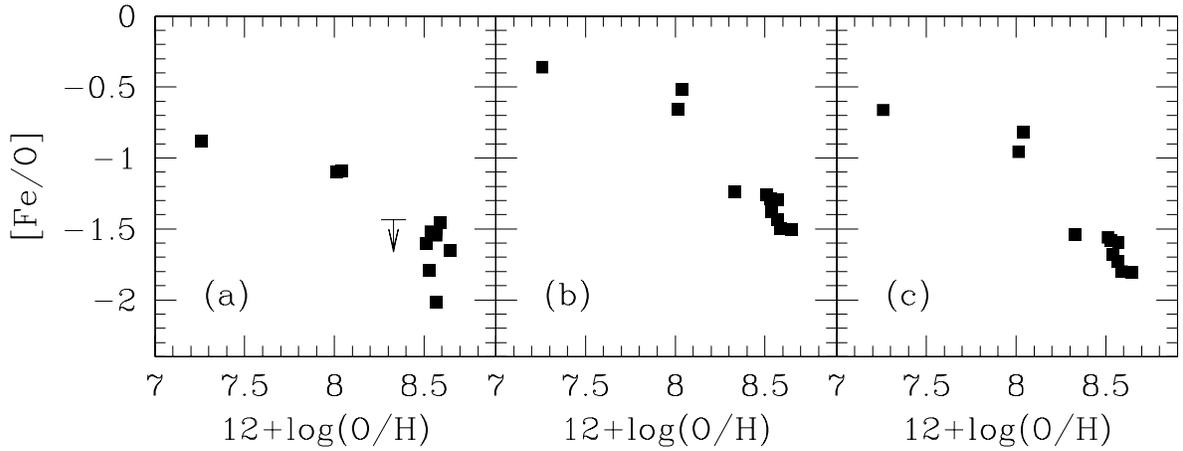}
\caption{Depletion factors
([Fe/O] $=\log(\mbox{Fe}/\mbox{O})-\log(\mbox{Fe}/\mbox{O})_\odot$) for our
sample objects as a function of their metallicity given by the O abundance.
Panel {\sl (a)} shows the depletion obtained assuming that both the Fe$^{++}$
and Fe$^{3+}$ collision strengths are approximately correct. Panel {\sl (b)}
shows the results when the Fe$^{++}$ collision strengths and the model predicted
concentrations are approximately correct. Panel {\sl (c)} is for the case where
the Fe$^{3+}$ collision strengths and the model predicted concentrations are
approximately correct but the Fe$^{++}$ collision strengths are too low by a
factor of $\sim2$, leading to Fe$^{++}$ abundances that are too high by a 
factor of 2.}
\label{dep}
\end{figure}

Since we are considering quite drastic changes in the atomic data involved in
the abundance calculation, we feel justified in using the Fe abundances implied
by the three possible explanations listed above as a way to constrain the true
Fe abundances in the gas of our sample objects.
Our set of Galactic \ion{H}{2} regions and PNe have Fe depletion
factors ($\log(\mbox{Fe}/\mbox{O})-\log(\mbox{Fe}/\mbox{O})_\odot$) below
$-1.3$; most of the Fe atoms are deposited onto dust grains in these nebulae.
Only a few per cent or less of their Fe atoms are in the gas phase.
The extragalactic \ion{H}{2} regions of our sample (LMC~30~Doradus, SMC~N88A,
and SBS~0335$-$052) show somewhat lower depletions and help define a trend of
increasing depletion with increasing metallicity (see Fig.~\ref{dep}).
The depletion factor in SBS~0335$-$052, one of the most metal-deficient galaxies
known, is only poorly constrained: it should be in the range $-0.9$ to $-0.4$.
The exact amount of depletion in this interesting object will not be known
until the [\ion{Fe}{4}] discrepancy is fully explained.

\acknowledgments
We thank Janet Simpson and Grazyna Stasi\'nska for helpful comments.
The comments of an anonymous referee also helped to improve the paper.
MR acknowledges support from Mexican CONACYT project J37680-E.\@
Support for RHR was from the NASA Long-Term Space Astrophysics (LTSA) Program
and the Astrophysics Data Program (ADP).
RHR thanks Scott McNealy for providing a Sun workstation.
The models were run on Cray computers at NASA/Ames (Consolidated Supercomputer 
Management Office) and at JPL; time on the latter was provided by funding 
from JPL Institutional Computing and Information Services and
the NASA Offices of Earth Science, Aeronautics, and Space Science.

\clearpage

\clearpage

\begin{deluxetable}{lllllc}
\tabletypesize{\small}
\tablecolumns{6}
\tablecaption{PHYSICAL CONDITIONS \label{t1}}
\tablewidth{0pt}
\tablehead{
\colhead{ } & \colhead{ } & \colhead{$N_e$} &
\colhead{$T_e$(low)\tablenotemark{a}} &
\colhead{$T_e$(high)\tablenotemark{a}} &
\colhead{}\\
\colhead{Object} & \colhead{ID} & \colhead{(cm$^{-3}$)} & \colhead{(K)} &
	\colhead{(K)} & \colhead{Reference}
}
\startdata
\objectname{30~Doradus} & LMC \ion{H}{2} & \phn\phn$440\pm190$ &
	$10800^{+350}_{-300}$ & $10000\pm200$  & 1\\
\objectname{IC~4846}  & Galactic PN & \phn$8700\pm3900$ &
	$12200^{+5000}_{-2000}$ & $10500^{+900}_{-600}$ & 2 \\
\objectname{M42} a & Galactic \ion{H}{2} & \phn$6400\pm2800$ &
	$10000^{+1600}_{-1000}$ & \phn$8300^{+600}_{-400}$ & 3 \\
\objectname{M42} b & Galactic \ion{H}{2} & \phn$8100\pm1600$ &
	\phn$9800^{+300}_{-200}$ & \phn$8300\pm100$ & 4 \\
\objectname{N88A bar}  & SMC \ion{H}{2} & $10200^{+4800}_{-6100}$ & $14200\pm400$ &
	$14200\pm400$ & 5 \\
\objectname{N88A sq.~A}  & SMC \ion{H}{2} & \phn1500$^{+4500}_{-1000}$ &
	$13500^{+900}_{-600}$ & $13500^{+900}_{-600}$ & 5 \\
\objectname{NGC~3576}  & Galactic \ion{H}{2} & \phn$3000\pm1200$ &
	\phn$8500\pm200$ & \phn$8500\pm100$ & 6 \\
\objectname{NGC~6210} &  Galactic PN & \phn$5900\pm2700$ &
	$11000^{+400}_{-500}$ & \phn$9600\pm200$ & 7 \\
\objectname{NGC~6826}  & Galactic PN & \phn$2300\pm700$ &
	$10600\pm300$ & \phn$9300^{+200}_{-100}$ & 7 \\
\objectname{NGC~6884} & Galactic PN  & $10200\pm3700$ &
	$11200^{+400}_{-300}$ & $11000^{+300}_{-200}$ & 7 \\
\objectname{SBS~0335$-$052} & \ion{H}{2} galaxy & \phn\phn300$^{+400}_{-270}$ &
	20200$^{+800}_{-700}$ & 20200$^{+800}_{-700}$ & 8 \\
\enddata
\tablenotetext{a}{$T_e$(low) is the $T_e$ derived from the
[\ion{N}{2}] diagnostic lines; it has been used to derive the O$^+$ and
Fe$^{++}$ abundances. $T_e$(high) is the $T_e$ implied by the
[\ion{O}{3}] diagnostic lines; it has been used to derive the O$^{++}$ and
Fe$^{3+}$ abundances. $T_e$[\ion{O}{3}] has been used to derive all ionic
abundances in N88A and SBS~0335$-$052 \citep{rod03}.}
\tablerefs{{Line intensities from (1) \citet{ape03}, (2) \citet*{hyu01},
(3) \citet{bal00}, (4) \citet{est04}, (5) \citet{kurt99}, (6) \citet{gar04}, (7)
\citet{liu04a}, (8) \citet{izo01}.}}
\end{deluxetable}

\clearpage

\begin{deluxetable}{lllllllll}
\tabletypesize{\small}
\rotate
\tablecolumns{9}
\tablecaption{IONIC AND TOTAL ABUNDANCES ($12+\log X$) \label{t2}}
\tablewidth{0pt}
\tablehead{
\colhead{Object} &
\colhead{$\mbox{O}^{+}/\mbox{H}^{+}$} & \colhead{$\mbox{O}^{++}/\mbox{H}^{+}$} &
\colhead{$\mbox{Fe}^{+}/\mbox{H}^{+}$} &
\colhead{$\mbox{Fe}^{++}/\mbox{H}^{+}$} &
\colhead{$\mbox{Fe}^{3+}/\mbox{H}^{+}$} & \colhead{O/H} &
\colhead{Fe/H\tablenotemark{a}} & \colhead{Fe/H\tablenotemark{b}}\\
\colhead{(1)} & \colhead{(2)} & \colhead{(3)} & \colhead{(4)} & \colhead{(5)} & 
\colhead{(6)} & \colhead{(7)} & \colhead{(8)} & \colhead{(9)}
}
\startdata
\objectname{30~Doradus} & $7.56^{+0.05}_{-0.06}$ & $8.25\pm0.04$ &
	\nodata & $5.22\pm0.07$ & $\le5.52$ &  $8.33\pm0.03$ & 
	$\le5.70$ & $5.90^{+0.06}_{-0.07}$\\
\objectname{IC~4846}  & $6.99^{+0.31}_{-0.55}$ &
	$8.50^{+0.09}_{-0.12}$ & \nodata & $4.70^{+0.24}_{-0.60}$ &
	$5.67^{+0.20}_{-0.33}$ & $8.51^{+0.09}_{-0.12}$ &
	$5.71^{+0.19}_{-0.29}$ & $6.05^{+0.24}_{-0.60}$\\
\objectname{M42} a  & $7.92^{+0.23}_{-0.32}$ & 
	$8.46^{+0.11}_{-0.15}$ & $4.73$: & $5.52^{+0.16}_{-0.20}$ &
	$5.46^{+0.17}_{-0.24}$ & $8.57^{+0.10}_{-0.13}$ &
	$5.83^{+0.12}_{-0.13}$ & $6.07^{+0.16}_{-0.20}$\\
\objectname{M42} b  & $7.88^{+0.06}_{-0.08}$ & 
	$8.43\pm0.03$ & $4.51$:\tablenotemark{c} & $5.39^{+0.06}_{-0.07}$ &
	$5.58\pm0.04$ & $8.54^{+0.02}_{-0.03}$ &
	$5.82\pm0.03$ & $5.96^{+0.06}_{-0.07}$\\
\objectname{N88A bar} & $6.96^{+0.11}_{-0.19}$ & $7.97\pm0.04$ &
	\nodata & $5.23\pm0.05$ & $5.55\pm0.06$ & $8.01\pm0.04$ &
	$5.72\pm0.04$ & $6.16\pm0.05$\\
\objectname{N88A sq.~A}  & $6.69^{+0.19}_{-0.14}$ & $8.02^{+0.06}_{-0.08}$ &
	\nodata & $5.12^{+0.11}_{-0.15}$ & $5.63^{+0.12}_{-0.16}$ &
	8.04$^{+0.06}_{-0.08}$ & 5.75$^{+0.09}_{-0.12}$ & $6.32^{+0.11}_{-0.15}$\\
\objectname{NGC~3576}  & $8.21^{+0.07}_{-0.08}$ & 
	$8.35\pm0.03$ & $4.54$:\tablenotemark{c} & $5.57^{+0.05}_{-0.06}$ &
	$5.65^{+0.18}_{-0.30}$ & $8.59\pm0.04$ &
	$5.93^{+0.11}_{-0.13}$ & $5.89^{+0.05}_{-0.06}$\\
\objectname{NGC~6210}  & $7.26^{+0.11}_{-0.12}$ & 
	$8.63\pm0.04$ & \nodata & $4.71\pm0.08$ &
	$5.76^{+0.18}_{-0.32}$ & $8.65\pm0.04$ &
	$5.79^{+0.17}_{-0.29}$ & $5.94\pm0.08$\\
\objectname{NGC~6826}  & $7.01\pm0.06$ & 
	$8.52^{+0.03}_{-0.04}$ & \nodata & $4.69\pm0.06$ &
	$5.47^{+0.18}_{-0.34}$ & $8.53^{+0.03}_{-0.04}$ &
	$5.54^{+0.16}_{-0.27}$ & $6.04\pm0.06$\\
\objectname{NGC~6884}  & $7.25^{+0.09}_{-0.12}$ & 
	$8.55^{+0.03}_{-0.04}$ & \nodata & $4.77^{+0.07}_{-0.08}$ &
	$5.23^{+0.18}_{-0.32}$ & $8.57^{+0.03}_{-0.04}$ &
	$5.36^{+0.14}_{-0.21}$ & $5.94^{+0.07}_{-0.08}$\\
\objectname{SBS~0335$-$052}  & $5.92^{+0.06}_{-0.07}$ & 
	$7.24\pm0.04$ & \nodata & $4.51\pm0.11$ &
	$5.07^{+0.18}_{-0.31}$ & $7.26^{+0.03}_{-0.04}$ &
	$5.18^{+0.15}_{-0.23}$  & $5.70\pm0.11$\\
\enddata
\tablenotetext{a}{Derived from the sum of the ionic abundances.}
\tablenotetext{b}{Derived from the Fe$^{++}$ abundance and the
$ICF$ scheme of equation (\ref{eq2}) implied by our photoionization models.}
\tablenotetext{c}{Derived using [\ion{Fe}{2}]~$\lambda$7155 (see
\citealt{gar04}).}
\end{deluxetable}
\clearpage


\begin{thebibliography}{}
\bibitem[Asplund et al.(2004)]{asp04} Asplund, M., Grevesse, N., Sauval, A. J.,
	Allende Prieto, C., \& Kiselman, D. 2004, \aap, 417, 751
\bibitem[Baldwin et al.(2000)]{bal00} Baldwin, J. A., Verner, E. M.,
	Verner, D. A., Ferland, G. J., Martin, P. G., Korista, K. T., \&
	Rubin, R. H. 2000, \apjs, 129, 229
\bibitem[Bautista et al.(1998)Bautista, Romano \& Pradhan]{baum98}
	Bautista, M. A., Romano, P., \& Pradhan, A.  K. 1998, \apjs, 118, 259
\bibitem[Bot et al.(2004)]{bot04} Bot, C., Boulanger, F., Lagache, G.,
	Cambr\'esy, L., \& Egret, D. 2004, \aap, 423, 567
\bibitem[Campbell et al.(1986)Campbell, Terlevich \& Melnick]{cam86} Campbell,
	A., Terlevich, R., \& Melnick, J. 1986, \mnras, 223, 811
\bibitem[Esteban et al.(2004)]{est04} Esteban, C., Peimbert, M.,
	Garc\'\i a-Rojas, J., Ruiz, M.-T., Peimbert, A., \& Rodr\'\i guez, M.
	2004, \mnras, 355, 229
\bibitem[Ferland(2003)]{fer03} Ferland, G. J. 2003, \araa, 41, 517
\bibitem[Froese Fischer \& Rubin(2004)]{fro04} Froese Fischer, C., \&
	Rubin, R. H. 2004, \mnras, 355, 461
\bibitem[Garc\'\i a-Rojas et al.(2004)]{gar04} Garc\'\i a-Rojas, J., Esteban,
	C., Peimbert, M., Rodr\'\i guez, M., Ruiz, M.-T., \& Peimbert, A. 2004,
	\apjs, 153, 501
\bibitem[Gruenwald \& Viegas(1992)]{gru92} Gruenwald, R. B., \& Viegas, S. M.
	1992, \apjs, 78, 153
\bibitem[Holweger(2001)]{hol01} Holweger, H. 2001, in AIP Conf. Proc. 598,
	Solar and Galactic Composition, ed. R. F. Wimmer-Schweingruber
	(New York: Springer-Verlag), 23 
\bibitem[Houck et al.(2004)]{hou04} Houck, J. R., et al.\ 2004, \apjl, 154, L211
\bibitem[Hyung et al.(2001)Hyung, Aller, \& Lee]{hyu01} Hyung, S., Aller, L. H.,
	\& Lee, W.-B. 2001, \pasp, 113, 1559
\bibitem[Izotov et al.(2001)Izotov, Chaffee, \& Schaerer]{izo01} Izotov, Y. I.,
	Chaffee, F. H., \& Schaerer, D. 2001, \aap, 378, L45
\bibitem[Izotov \& Thuan(1999)]{izo99} Izotov, Y. I., \& Thuan, T. X. 1999,
	\apj, 511, 639
\bibitem[Jenkins(2004)]{jen04} Jenkins, E. B. 2004, in Carnegie Observatories
	Astrophysics Series 4, Origin and Evolution of the Elements, ed. A.
	McWilliam \& M. Rauch (Cambridge: Cambridge Univ. Press), 339
\bibitem[Kingdon \& Ferland(1996)]{kin96} Kingdon, J. B., \& Ferland, G. J.
	1996, \apjs, 106, 205
\bibitem[Kjeldsen et al.(2002a)]{kj02a} Kjeldsen, H., Kristensen, B., Brooks,
	R. L., Folkman, F., Knudsen, H., \& Andersen, T. 2002a, \apjs, 138, 219
\bibitem[Kjeldsen et al.(2002b)]{kj02b} Kjeldsen, H., Kristensen, B., Folkman,
	F., \& Andersen, T. 2002b, J. Phys. B, 35, 3655
\bibitem[Kurt et al.(1999)]{kurt99} Kurt, C. M., Dufour, R. J., Garnett, D. R.,
	Skillman, E. D., Mathis, J. S., Peimbert, M., Torres-Peimbert. S., \&
	Ruiz, M.-T. 1999, \apj, 518, 246
\bibitem[Liu et al.(2004a)]{liu04a} Liu, Y., Liu, X.-W., Luo, S.-G., \& Barlow, M.
	J. 2004a, \mnras, 353, 1231
\bibitem[Liu et al.(2004b)]{liu04b} Liu, Y., Liu, X.-W., Barlow, M., \& Luo, S.-G.
	J. 2004b, \mnras, 353, 1251
\bibitem[McLaughlin et al.(2002)]{mcl02} McLaughlin, B.M., Scott, M.P.,
	Sunderland, A.G., Noble, C.J., Burke, V.M., \& Burke, P.G.\ 2002,
	J.\ Phys.\ B, 35, 2755
\bibitem[Mel\'endez(2004)]{mel04} Mel\'endez, J. 2004, \apj, 615, 1042
\bibitem[Moore et al.(2004)Moore, Hester, \& Dufour]{moo04} Moore, B. D.,
	Hester, J. J., \& Dufour, R. J. 2004, \aj, 127, 3484
\bibitem[Nahar(1996a)]{nah96a} Nahar, S. N. 1996a, \pra, 53, 1545
\bibitem[Nahar(1996b)]{nah96b} Nahar, S. N. 1996b, \pra, 53, 2417
\bibitem[Nahar(1997)]{nah97} Nahar, S. N. 1997, \pra, 55, 1980
\bibitem[Nahar(1999)]{nah99} Nahar, S. N. 1999, \apjs, 120, 131
\bibitem[Nahar \& Pradhan(1994)]{nah94} Nahar, S. N., \& Pradhan, A. K. 1994, J.
	Phys. B, 27, 429 
\bibitem[Osterbrock(1989)]{ost89} Osterbrock, D. E. 1989, Astrophysics of
	Gaseous Nebulae and Active Galactic Nuclei (Mill Valley: University
	Science Books)
\bibitem[Peimbert(2003)]{ape03} Peimbert, A. 2003, \apj, 584, 735
\bibitem[Roche et al.(1987)Roche, Aitken, \& Smith]{roc87} Roche, P. F., Aitken,
	D. K., \& Smith, C. H. 1987, \mnras, 228, 269
\bibitem[Rodr\'\i guez(1996)]{rod96} Rodr\'\i guez, M. 1996, \aap, 313, L5
\bibitem[Rodr\'\i guez(1999)]{rod99} Rodr\'\i guez, M. 1999, \aap, 348, 222
\bibitem[Rodr\'\i guez(2002)]{rod02} Rodr\'\i guez, M. 2002, \aap, 389, 556
\bibitem[Rodr\'\i guez(2003)]{rod03} Rodr\'\i guez, M. 2003, \apj, 590, 296
\bibitem[Rubin et al.(1991a)]{rub91a} Rubin, R. H., Simpson, J. P., Haas, M. R.,
	\& Erickson, E. F. 1991a, \apj, 374, 564
\bibitem[Rubin et al.(1991b)]{rub91b} Rubin, R. H., Simpson, J. P., Haas, M. R.,
	\& Erickson, E. F. 1991b, \pasp, 103, 834
\bibitem[Rubin et al.(1997)]{rub97} Rubin, R. H., et al.\ 1997, \apjl, 474, L131
\bibitem[Savage \& Sembach(1996)]{sav96} Savage, B. D., \& Sembach, K. R. 1996,
	\araa, 34, 279
\bibitem[Shetrone et al.(2001)Shetrone, C\^ot\'e, \& Sargent]{she01} Shetrone,
	M. D., C\^ot\'e, P., \& Sargent, W. L. W. 2001, \apj, 548, 592
\bibitem[Simpson et al.(2004)]{sim04} Simpson, J. P., Rubin, R. H., Colgan, S.
	W. J., Erickson, E. F., \& Haas, M. R. 2004, \apj, 611, 338
\bibitem[Sofia et al.(1994)Sofia, Cardelli, \& Savage]{sof94} Sofia, U. J.,
	Cardelli, J. A., \& Savage, B. D. 1994, \apj, 430, 650
\bibitem[Stasi\'nska(1990)]{sta90} Stasi\'nska, G. 1990, \aaps, 83, 501
\bibitem[Sternberg et al.(2003)Sternberg, Hoffmann, \& Pauldrach]{stern03}
	Sternberg, A., Hoffmann, T. L., \& Pauldrach, A. W. A. 2003, \apj, 599,
	1333
\bibitem[Venn et al.(2001)]{ven01} Venn, K. A., Lennon, D. J., Kaufer, A.,
	McCarthy, J. K., Przybilla, N., Kudritzki, R. P., Lemke, M., Skillman,
	E. D., \& Smartt, S. J. 2001, \apj, 547, 765
\bibitem[Verner et al.(1996)]{ver96} Verner, D. A., Ferland, G. J., Korista,
	K. T., \& Yakolev, D. G. 1996, \apj, 465 487
\bibitem[Verner et al.(2000)]{ver00} Verner, E. M., Verner, D. A., Baldwin, J.
	A., Ferland, G. J., \& Martin, P. G. 2000, \apj, 543, 831
\bibitem[Vladilo(2004)]{vla04} Vladilo, G. 2004, \aap, 421, 479
\bibitem[Welty et al.(1999)]{wel99} Welty, D. E., Frisch, P. C., Sonneborn, G.,
	\& York, D. G. 1999, \apj, 512, 636
\bibitem[Welty et al.(2001)]{wel01} Welty, D. E., Lauroesch, J. T., Blades, C.,
	Hobbs, L. M., \& York, D. G. 2001, \apjl, 554, L75
\bibitem[Whittet(2003)]{whi03} Whittet, D. C. B. 2003, Dust in the Galactic
	Environment (Bristol: IOP Publishing)
\bibitem[Zhang(1996)]{zha96} Zhang, H. L. 1996, A\&AS, 119, 523
\end{thebibliography}
\end{document}